\documentclass[onecolumn,12pt]{aastex62}

\usepackage{blindtext}
\usepackage[colorinlistoftodos]{todonotes}
\usepackage{lineno}
\usepackage{amsmath}
\usepackage{MnSymbol}
\usepackage{graphicx}

\begin{document}

\title{Which countries are leading high-impact science in astronomy?}

\author{Juan P. Madrid} 

\affiliation{Department of Physics and Astronomy, The University of Texas Rio Grande Valley, Brownsville, TX 78520, USA}

\begin{abstract}

Recent news reports claim that China is overtaking the United States and all other countries in 
scientific productivity and scientific impact. A straightforward analysis of high-impact papers 
in astronomy reveals that this is not true in our field. In fact, the United States continues to 
host, by a large margin, the authors that lead high-impact papers. Moreover, this analysis shows
that 90\% of all high-impact papers in astronomy are led by authors based in North America and 
Europe. That is, only about 10\% of countries in the world  host astronomers
that publish ``astronomy's greatest hits".\\

\end{abstract}

\section{Introduction}

Recent news reports claim that China is overtaking the United States in scientific 
research output and in particular high-impact papers\footnote{Japan falls out of top 
10 nations with most-cited scientific papers Fujinami, Y. The Asahi Shimbun (August 2022) 
\textcolor{blue}{\href{https://www.asahi.com/ajw/articles/14691980}{https://www.asahi.com/ajw/articles/14691980}}}. Several news sources quote a 
report from the National Institute of Science and Technology of Japan that found 
that in recent years, China has been publishing not only the highest number of 
scientific research papers, but also the most cited ones, overtaking all other 
countries including the United States\footnote{China overtakes the US in scientific 
research output. Lu, D., The Guardian (August 2022) \textcolor{blue}{\href{https://www.theguardian.com/world/2022/aug/11/china-overtakes-the-us-in-scientific-research-output}
{https://www.theguardian.com/world/2022/aug/11/china-overtakes-the-us-in-scientific-research-output}}}. 
Is this true for astronomy?\\

\section{Methods}

In the NASA Astrophysics Data System (ADS) our field has a pioneering  digital library 
\citep{kurtz2000, accomazzi2015}. The ADS gives us access to a wide range of bibliographic information with just 
a few clicks. The capabilities of the ADS have been routinely used to evaluate the productivity and impact 
of individual astronomical observatories, e.g.\ Lick \citep{smith2020}, Spitzer \citep{scire2022},
Chandra \citep{blecksmith2005}, and HST \citep{meylan2004}.

A simple ADS query, resolved in a fraction of a second, will reveal the 
most cited papers for any given year. Using the ADS, we ranked  all 
papers published in the year 2020 by the number of citations. Why 2020? 
We must wait for at least a year for papers to gather citations but we also want 
to have the most recent values \citep{meylan2004}. 

We kept tabs on the author's affiliation for the 
first author, or the corresponding author, for these 100 most cited astronomy papers, 
published in 2020. These are the papers that can be considered to have the most influence 
in the field, or ``Astronomy's Greatest Hits" \citep{frogel2010}. Crediting papers to the 
host country of the first author has been used as a straightforward method to estimate 
the impact of astronomical research from different countries \citep{sanchez2004}.

Publications with the highest impact accumulate a large number of citations within the 
first year. Indeed, a limited number of papers swiftly accumulate a very large 
quantity of citations. The 100 most cited papers (published in 2020) constitute only 
0.1\% of the total records indexed by the ADS for that year but represent 11\% of 
the total citations. For those mathematically inclined, the distribution of citations
for papers published in a given year, is well described by a declining exponential
\citep{madrid2006}.

Each of these 100 high-impact papers count for 1\% of the total in Table 1. 
What if a paper has two corresponding authors from two different countries? We 
allocated 0.5\% for each country, which explains the fractional values.\\
\bigskip

\section{Results}

Scientists based in the United States published the largest share of high-impact papers
in 2020, that is, 36\% of the total. The United States is followed by Germany and the United
Kingdom both publishing 9.5\% of the total. If all countries belonging to the the European Union 
are added together they would represent 36\% of the total of high-impact papers (excluding the UK).
European countries that do not appear in Table 1 but that have a contribution to the most cited 
papers are: Belgium (2\%), Denmark, Hungary, Poland, and Sweden all with 1\% of the total, 
and Czechia with 0.5\%.

The most influential work in astronomy is overwhelmingly led by scientists based in the ``North Atlantic".
That is, 90\% of all high-impact papers in Astronomy are led by authors based in North America and 
Europe. Contrary to recent reports, China only leads 2.5\% of all high-impact papers for the year we 
evaluated. This work also highlights the disparity across the world when it comes to game-changing
scientific work: only 13\% of all countries in the world (25 out 193) host scientists that pilot 
the top 100 high-impact papers. 

For the year 2020, the statistics of high-impact papers were shaped by the results of the Planck
mission and the LIGO-Virgo collaboration. The cosmological parameters published by the Planck
collaboration, at almost 8,000 citations, is the most cited paper for that year \citep{planck2020}.\\

{\large
\begin{table}[h!]
\begin{center}
\begin{tabular}{llr} 
\multicolumn{3}{c}{Table 1. High-Impact Papers by Country}\\
\hline
Rank & Country        &    \%\\
\hline
1  &  USA             &  36.0\% \\
2  &  Germany         &  9.5\%  \\
3  &  UK              &  9.5\%  \\
4  &  Italy           &  7.6\%  \\
5  &  France          &  4.2\%  \\
6  &  Canada          &  4.0\%  \\
7  &  The Netherlands &  3.5\%  \\
8  &  Switzerland     &  3.0\%  \\
9  &  Spain           &  2.7\%  \\
10 &  China           &  2.5\%  \\
\hline
\end{tabular}
\end{center}
\end{table}
}



\bigskip

\begin{thebibliography}{}

\bibitem[Accomazzi et al.(2015)]{accomazzi2015} Accomazzi, A., Kurtz, M.\ J., Henneken, E.\ A. et al.\ 2015, ASP Conference Series, Vol. 492
San Francisco: Astronomical Society of the Pacific, p.189

\bibitem[Blecksmith et al.(2005)]{blecksmith2005} Blecksmith, S., Bright, J., Rots, A.\ H., et al.\ 2005, adass XIV, 347, 380

\bibitem[S\'anchez \& Benn(2004)]{sanchez2004} S\'anchez, S.\ F.\ \& Benn, C.\ R.\ 2004, Astron.\ Nachr.\ 325, 445

\bibitem[Frogel(2010)]{frogel2010} Frogel, J.\ A.\ 2010, PASP, 122, 1214

\bibitem[Kurtz et al.(2000)]{kurtz2000}  Kurtz, M.\ J.,  Eichhorn, G., Accomazzi, A., Grant, C.\ S.,
Murray, S. S., Watson, J.\ M.\ 2000, A\&AS, 143, 41

\bibitem[Madrid et al.(2006)]{madrid2006} Madrid, J.\ P. \& Macchetto, F.\ D.\ 2006, Bulletin of the American 
Astronomical Society, Vol.\ 38, p.\ 1286

\bibitem[Meylan et al.(2004)]{meylan2004} Meylan, G., Madrid, J.\ P., \& Macchetto, D.\ 2004, PASP, 116, 790

\bibitem[Planck collaboration(2020)]{planck2020} Planck collaboration, 2020, A\&A, 641, 6

\bibitem[Scire et al.(2022)]{scire2022} Scire, E., Rebull, L., \&  Laine, S.\ 2022, PASP,  134, 055001

\bibitem[Smith \& Shetrone(2020)]{smith2020}  Smith, G.\ H.\ \& Shetrone, M.\ 2020, PASP, 132, 125002


\end{thebibliography}
\end{document}